\documentclass[a4paper,11pt]{article}
\pdfoutput=1 

\usepackage{jheppub} 

\usepackage[T1]{fontenc}

\usepackage[applemac]{inputenc} 

\newcommand{\be}{\begin{equation}}
\newcommand{\ee}{\end{equation}}
\newcommand{\bea}{\begin{eqnarray}}
\newcommand{\eea}{\end{eqnarray}}
\newcommand{\non}{\nonumber}

\newcommand{\ra}{\rightarrow}
\newcommand{\al}{\alpha}

\title{Explosive particle production in non-commutative inflation}

\author[a]{Hideki Perrier,}
\author[a]{Ruth Durrer,}
\author[b]{and Massimiliano Rinaldi.}

\affiliation[a]{Universit\'e de Gen\`eve, D\'epartement de Physique Th\'eorique and CAP,\\
24 quai Ernest-Ansermet, CH-1211 Gen\`eve 4, Switzerland.}
\affiliation[b]{Namur Center for Complex systems (naXys),
University of Namur,\\ Rempart de la Vierge 8, 5000 B - Namur, Belgium.}

\emailAdd{hideki.perrier@unige.ch}
\emailAdd{ruth.durrer@unige.ch}
\emailAdd{mrinaldi@fundp.ac.be}

\abstract{We consider a model of inflation which has recently been proposed in the literature and where inflation is induced by corrections to the energy density coming from the non-commutativity of spacetime. We show that the very rapid inflationary expansion typical of this model is responsible for a burst of particle production which ends inflation and leads to a radiation-dominated phase. We analytically estimate the energy density of these particles and we  confront the results with more precise numerical calculations. We  estimate the number of inflationary e-folds before the back-reaction of the radiation energy density overcomes the non-commutative effects and terminates inflation naturally. }

\begin{document}
\maketitle
\flushbottom

\section{Introduction}

The inflationary paradigm is based on the assumption that the early Universe has experienced a phase of 
quasi-exponential expansion, which is usually driven by the dynamics of a minimally coupled scalar field \cite{infl,book1,book2}. The potential associated to the scalar field and/or the initial condition must be carefully modeled so that the 
inflationary expansion lasts sufficiently long to solve the horizon and flatness problems and, at the end of inflation, the scalar field can decay into a sufficient amount of matter. Even though simple single field inflation is in good agreement with observations~\cite{Komatsu}, it remains a phenomenological model as the origin of the scalar field is unknown. Another serious problem is the large degree of extrapolation of physical laws, that are extended to the extreme, near-Planckian regime at which many models of inflation take place.

It is a well-known fact that General Relativity (GR), as it stands, no longer holds when the energy density reaches Planckian values. Several attempts to unify GR and quantum mechanics, in order to find a consistent quantum theory of gravity, are under investigation. Among these, there is one line of research that suggests the extension of the Heisenberg uncertainty principle to pairs of coordinates. In this theory, the quantum algebra relating the position and the momentum of a single particle is implemented by a new non-commutative relation of the form $[x_{i},x_{j}]=i\theta_{ij}$, where $\theta_{ij}$ is of the order of the Planck length squared~\cite{ncfund}. This commutator clearly restricts non-commutative effects to very special situations, where the energy density is of the order of the Planck value. Such extreme conditions can occur near singularities and at the onset of an inflationary phase. Non-commutative effects in these situations have been already investigated, see e.\ g.\ \cite{BraMag}-\cite{Banerjee}.

In this paper we consider a model of non-commutative inflation proposed in Ref.~\cite{maxinf}, which is based on the so-called coherent state formalism developed in Refs.~\cite{sma}. The model predicts a rapid phase on inflation driven uniquely by non commutative effects, that act on the quantum spacetime fluctuations, modeled as a scalar field, in such a way that they behave as a time-dependent dark energy term with a Gaussian density profile. 

The advantage of this model is that it does not require any ad hoc potential and the inflationary phase ends naturally when the energy density drops below a certain value. In the simplest setup, any form of pre-existing matter is absent and the Universe is initially empty  at $t\ra -\infty$. In this paper we study particle creation in this time-dependent background at high curvature.  The principle of this  effect was studied long ago, and it is known to be quite effective in a quasi-de Sitter space \cite{ParProd}-\cite{full}. In the standard inflationary scenario, the production of hot and relativistic particles is subdominant during inflation and it occurs only after the slow roll phase, when the inflaton starts oscillating, at the onset of a  re-heating or pre-heating phase.

In our model there is no inflaton therefore quantum particle production is the only way to fill the Universe with radiation. In fact, as we will show, the non-commutative expansion is so rapid that particle creation is actually ``explosive'' and it eventually ends inflation by its back-reaction on the spacetime. In this paper we study this process in detail and calculate the energy density of the generated particles. 

In Sec.~\ref{ncsec} we briefly review the non-commutative inflationary model at hand and in Sec.~\ref{ppsec} we give an analytic estimate of the particle production. In Sec.~\ref{nrsec} we present the numerical results, and we conclude in Sec.~\ref{consec} with some remarks.


\section{Non-commutative inflation}\label{ncsec}


In Ref.\ \cite{maxinf} one of us has shown that, in a homogeneous and isotropic Universe,  non-commutative effects lead to a dynamical vacuum energy density that has the form
\bea\label{endens}
\langle\rho\rangle_{\rm NC}\simeq \rho_{0}\exp (-\Delta t^{2}/\theta),
\eea
where $\theta$ has the dimension of a  length squared. In static spacetimes, a similar formula holds for the energy density for the mass at the center of a black hole, see \cite{Nico}.  As shown in \cite{maxinf}, these results follow from a redefinition of relativistic quantum fields and the value of $\theta$ turns out to be of the order of the Planck length squared $\ell_{p}^{2}$. Remarkably, the last statement is consistent with traditional vacuum energy calculations. In fact, a simple argument shows that the expectation value for the vacuum energy can be calculated with the integral
\bea
\langle\rho\rangle=\int_{0}^{\Lambda}dk{4\pi k^{2}\over (2\pi)^{3}}{1\over 2}\sqrt{k^{2}+m^{2}},
\eea
which yields, in the limit $\Lambda\gg m$, $\langle\rho\rangle=\Lambda^{4}/(16\pi^{2})$. The ultraviolet cutoff $\Lambda$ identifies the transplanckian frontier and it can be set to $\Lambda=(8\pi G)^{-1/2}$, where $G$ is the Newton constant (with $\hbar=c=1$), see e.g. \cite{weimbergrev,Maggiore}. 

In the coherent state non-commutative approach, the vacuum energy for the fields should be computed instead as \cite{maxlast}
\bea\label{NCrho}
\langle\rho\rangle_{\rm NC}=\int_{0}^{\infty}dk{4\pi k^{2}\,e^{-2\theta k^{2}}\over (2\pi)^{3}}{1\over 2}\sqrt{k^{2}+m^{2}},
\eea
where now $\theta$ is the non-commutative parameter that corresponds to a length squared. In the limit where $m$ is negligible, this integral gives $\langle\rho\rangle_{\rm NC}=1/(8\theta^{2})$. By equating the two results, one finds that
\bea\label{thetaest}
\sqrt{\theta}=(128)^{1/4}\pi\ell_{p}\simeq 10.6 \, \ell_{p},
\eea
where $\ell_{p}=\sqrt{G}$ is the Planck length. The two calculations are then consistent provided the parameter $\sqrt{\theta}$ is of the order of the Planck length. To account for this result,  we  will
parameterize $\theta$ by
\be
\sqrt{\theta} =\al\ell_p \,.
\ee
Naturally, we expect $\al$ to be a constant of order unity. In  the derivation given in Ref.~\cite{maxinf} it was found that the leading term of the deWitt--Schwinger expansion of the energy momentum tensor for a single massless scalar field in the non-commutative field theory, based on the coherent state formalism, is
\be\label{1scalarf}
\langle T_{\mu\nu} \rangle \simeq \frac{1}{32\pi^2\theta^2}g_{\mu\nu} \,,
\ee
which corresponds to an energy density somewhat smaller than the above estimation. However, there is the possibility that the vacuum is determined by several fields increasing the value of its energy density.

The formula \eqref{endens} yields the Friedmann equation
\bea\label{NCFried}
\left( a'\over a\right)^2={8\pi G\over 3}\rho_{\rm nc}(t)\equiv H_m^2\, e^{-(t-t_{\max})^2/4\theta}\ ,
\eea
where $a$ is the scale factor of the metric $ds^2=-dt^2+a^2(t)\delta_{ij}dx^idx^j$ and $H=a'/a$. From now on, the prime indicates the derivative with respect to the physical time $t$ while a dot with respect to the conformal time $\eta$, defined by $dt=ad\eta$. Without loss of generality, we can also set $t_{\max}=0$. The parameter $H_{m}$ absorbs other eventual factors, such as the number of fundamental fields that are excited by non-commutative effects, and it is time-independent. If we equate the maximum energy density, $\rho_{\max}=\rho(0)$, with the $\langle T^{0}_{\,\,\,\,0}\rangle$ component of the expression  (\ref{1scalarf}) we find
\be \label{1scalarHtheta}
 H_m^2 =\frac {\ell_p^2}{12\pi\theta^2} = \beta^2\ell_p^{-2} ~\mbox{ with }~ 
 \beta = \frac{1}{\al^2\sqrt{12\pi}} \,.
\ee
In the following we shall, however, relax this condition on $H_m$ and let $\beta$ be arbitrary. As mentioned above, a larger
value for $\beta$  is easily justified by allowing for $N$ massless scalar fields
contribution to $\rho_{\max}$, such that $\beta^2 \ra N\beta^2$.

Like in the black hole case, Eq.~(\ref{NCFried}) has to be interpreted as the impossibility of localizing an arbitrary amount of energy at the time $t=0$ as the result, for example, of a collapse. However, to simplify our calculations, we choose initial conditions supposing that the Universe starts in the far past  empty and with a small value of $a$ and of the energy density. Thus, we only consider the non-commutative energy density and neglect any other contribution from radiation or cosmological constant. This is one of the advantages of this model: we can let the system evolve from very simple vacuum initial conditions. The only arbitrary choice is the sign of $H$, which we take to be positive. 

\begin{figure}[tbp]
\centering
\includegraphics[width=0.55\textwidth]{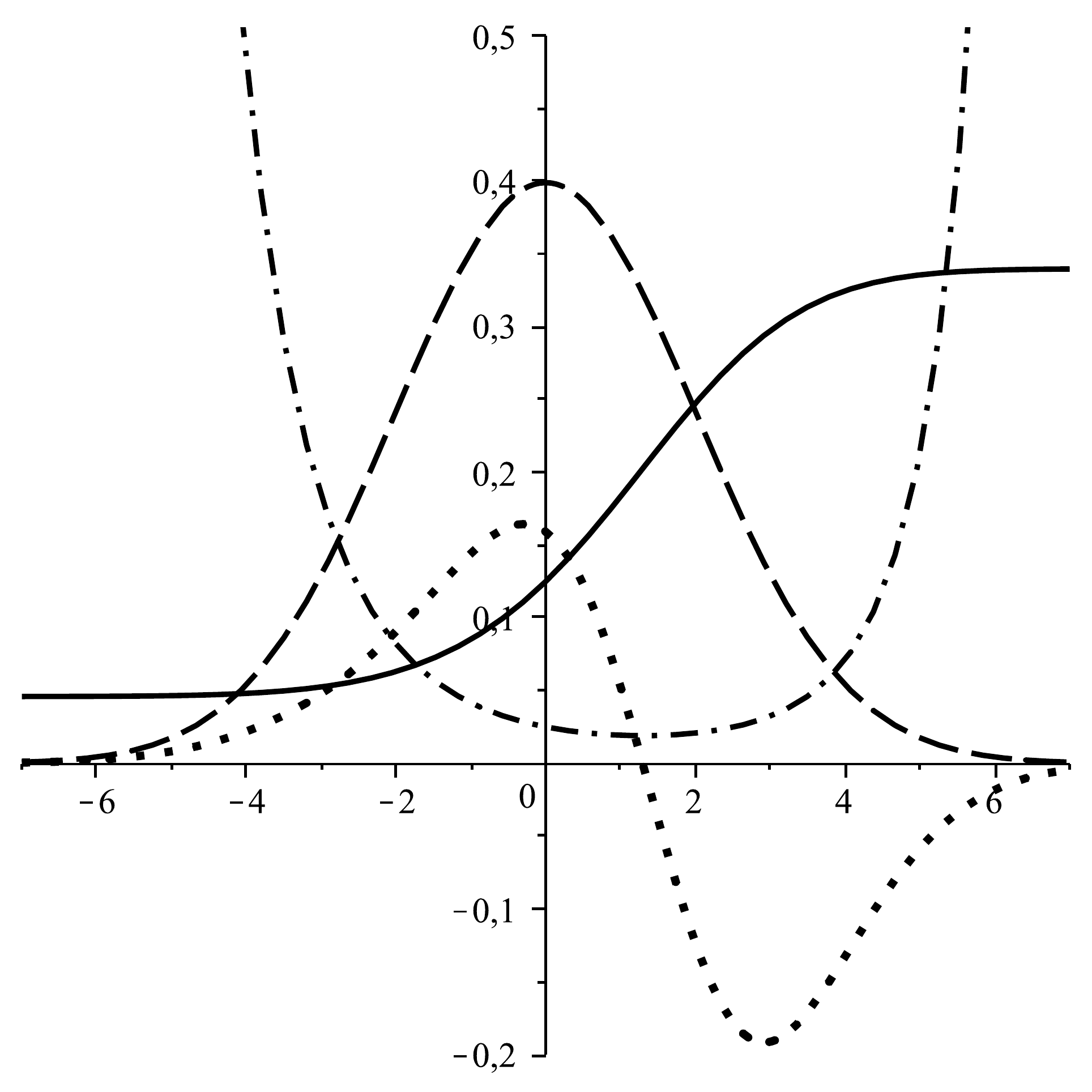}
\hfill 
\caption{\label{FigA} Qualitative behavior of $a$ (solid line), $a''$ (dotted line), $H$ (dashed line), and $(aH)^{-1}$ (dot-dashed line) as functions of  time.}
\end{figure}

With these hypothesis, we easily find the solution to Eq.~(\ref{NCFried}), which reads
\bea\label{scaleF}
a(t)&=&a_m\exp\left[{H_m\sqrt{2\pi\theta}}\,\, {\rm Erf}\left(t\over 2\sqrt{2\theta}\right)\right] =
a_m\exp\left[{\sqrt{2\pi}\beta\al}\,\, {\rm Erf}\left(t\over 2\sqrt{2\theta}\right)\right] 
\ ,\\   && {\rm Erf}(x')= {2\over\sqrt{\pi}}\int_0^{x'}\,e^{-x^2}dx\ , \nonumber
\eea
where $a_m$ is an integration constant. It can be checked that $a''$ is initially positive and then changes sign at the time when the comoving Hubble length $(aH)^{-1}$ reaches its minimum, while $H$ reaches its maximum, $H_m$, at $t=0$, see fig.~\ref{FigA}.  In other words, the global evolution of the scale factor is very similar to the case of the pre-Big Bang scenario of string cosmology~\cite{VenGa}. $H'(t)=-{t\over 4\theta}H(t)$ is positive for $t<0$ and negative for $t>0$.

The number of e-foldings of inflation can be estimated as
\be\label{e:efolds} 
N_e  \simeq \ln\left(a(t\ra\infty)\over a(t\ra -\infty)\right) = \sqrt{8\pi H^2_m\theta} = \sqrt{8\pi}\al\beta\,.
\ee
If we insert $\beta$ from (\ref{1scalarHtheta}), requiring 60 e-folds as in standard inflation, 
yields $\al\simeq 0.014$. 
Leaving both, $\al$ and $\beta$ free, the requirement of at least 60 e-folds of  inflation implies the relation
\be
\alpha\beta>\frac{60}{\sqrt{8\pi}} \simeq 12\ .
\ee
Note that this is the 'naively estimated' number of e-folds. In Sec.~\ref{nrsec} we shall show that the true number of e-folds is reduced by up to a factor of 2 due to backreaction from particle creation.

By imposing the second Friedmann equation, $\rho'+3H(\rho+p)=0$, we find the effective pressure $p$. 
Defining the equation of state parameter $\omega$ by  $p(t)=\omega(t)\rho(t)$, we obtain
\bea
\omega= -1-\frac{2}{3}\frac{H'}{H^2} \simeq -1+{t\over{6\theta H_m}}+{\cal O}\left({t^3\over\theta^2}\right)\ .
\eea
This leads to a dynamical crossing of the value $\omega=-1$ value at $t=0$ where $H(t)$ reaches its maximum.  At negative times, this models exhibits an effective  ``phantom'' behavior, $\omega<-1$.


\section{Particle production}\label{ppsec}


It is a long-known fact that a sufficiently rapid growth of the scale factor leads to particle 
production~\cite{ParProd}-\cite{full}. Technically, a meaningful calculation requires two asymptotic regimes where the the spacetime is flat, or slowly varying, so that the particle number is a well-defined quantity. In our simple model the scale factor is asymptotically constant both in the past and in the future, an ideal situation to calculate the energy density of the produced particles. 

The mode equation cannot be solved analytically because of the complicated form of the scale factor  (\ref{scaleF}), so we need to resort to numerical approximations.
However, some analytic estimates can be given. 

Let us consider the action for a massless scalar field 
\bea
S=-{1\over 2}\int d^{4}x\sqrt{-g}\,g^{\mu\nu}\partial_{\mu}\phi\partial_{\nu}\phi.
\eea
On a Friedmann-Lema\^\i tre (FL)  background with metric $ds^{2}=a(\eta)^{2}(-d\eta^{2}+d\vec{x}^{2})$ in conformal time, it becomes
\bea
S=\int d^{3}xd\eta\left[{1\over 2}a^{2}\left( \dot\phi^{2}-(\nabla\phi)^{2}\right)\right].
\label{action}
\eea
Setting $\chi(\eta)=a(\eta)\phi(\eta)$ this action takes the canonical form
\begin{equation}\label{chiaction}
S=\frac{1}{2}\int d^3\textbf{x} d\eta \,  \left( \dot{\chi}^2 - \left(\nabla\chi\right)^2  + 
   \frac{\ddot{a}}{a}\chi^2 \right)\, ,
\end{equation}
 and the associated equation of motion, in Fourier space, reads
\be\label{chimotion}
\ddot\chi_{k}+\omega_{k}^{2}(\eta)\chi_{k}=0, \quad
\mbox {where } \quad \omega_{k}^{2}(\eta)=k^{2}-\ddot a/a=k^2+m_\phi^2(\eta)\,. 
\ee
At early times $m_\phi^2(\eta)<0$ indicating an instability and at late times it becomes positive. 
Following the usual quantization procedure, the field $\chi$ is promoted to an operator $\hat \chi_{k}(\eta)=\hat a_{k}v_{k}^{*}(\eta)+\hat a_{-k}^{\dagger}v_{k}(\eta)$ and the modes $v$ satisfy the classical equation of motion
\bea\label{vmodes}
\ddot v_{k}+\omega_{k}^{2}(\eta)v_{k}=0.
\eea
In our case, we find it more convenient to work in physical time, since the form of the scale factor and its derivatives are given explicitely in $t$. The above equation then takes the form 
\bea\label{phystime}
v''+\left(a'\over a\right)v'+\left[\left(k\over a\right)^{2}-\left(a''\over a\right)-\left(a'\over a\right)^{2}\right]v=0.
\eea
In terms of the dimensionless time variable $T=t/\sqrt{8\theta}$, momentum 
$K=k\sqrt{8\theta}$, and the dimensionless mode function $V=v/(8\theta)^{1/4}$ the mode equation finally reads
\bea
{d^{2}V\over dT^{2}}+{1\over a}{da\over dT}{dV\over dT}+\left[ \left(K\over a\right)^{2}-{1\over a}\left(d^2a\over dT^{2}\right)-{1\over a^{2}}\left(da\over dT\right)^{2}\right]V=0,
\eea
and the initial vacuum state, corresponding to $a=$ constant,  is $V=\exp(-iKT)/\sqrt{2K}$. In the 
variable $T$, the scale factor and the background energy density are  given by
\bea
a(T) &=& a_m\exp\left[\frac{N_e}{2} \;{\rm Erf}(T)\right]\ ,  \label{e:scal}\\
\rho_{\rm nc}(T) &=& \rho_{\max}\exp(-2T^2)  \label{e:rnc},\quad
\rho_{\max} = \frac{3}{8\pi}\frac{\al^4\beta^2}{\theta^2} =\frac{3}{64\pi^2}\frac{N_e^2\al^2}{\theta^2}\,.
\eea
We also report the expression of the  dimensionless dynamical mass $M_{\phi}(\eta)$ defined in Eq.\ \eqref{chimotion}, namely
\bea
8\theta m_\phi^2(T)\equiv M_\phi^2(T)  &=& -\frac{a^2_mN_e^2}{\pi}\exp\Big[N_e \;{\rm Erf}(T)\Big]
\left[2-\frac{\sqrt{\pi}T}{N_e} \exp (T^2)\right]\, .
\eea


\subsection{Definitions of the energy density}
The usual definition of the energy-momentum tensor in GR is
\begin{equation}
T_{\mu \nu}=-\frac{2}{\sqrt{|g|}}\frac{\delta( \sqrt{|g|}\mathcal{L}_M)}{\delta g^{\mu \nu}}\,,
\label{gravT}
\end{equation}
where $\mathcal{L}_M$ is the matter Lagrangian density. 
This is the definition of the energy-momentum density that has to be used in the Einstein equations. 
 In general, it does not agree with the \textit{canonical} energy-momentum tensor obtained via the Noether theorem, but differs from it only by a total derivative.

The Hamiltonian related to the action \eqref{action} is
\begin{equation}
H_{\phi}=\int d^3 \textbf{x} \,a^4\left(\frac{1}{2a^2}\dot{\phi}^2 +\frac{1}{2a^2} \left(\nabla \phi\right)^2 \right).
\label{phiH}
\end{equation}
In the previous sections, we found useful to change the variable $\phi \rightarrow \chi=a\phi$, which is a canonical transformation, in order to obtain the standard equation of motion \eqref{chimotion}. The action for $\chi$ given in eq.~(\ref{chiaction}) leads to the Hamiltonian
\begin{equation}
H_{\chi}=   \frac{1}{2}\int d^3 \textbf{x} \left( \dot{\chi}^2 + \left(\nabla \chi\right)^2 -
       \frac{\ddot{a}}{a}\chi^2\right)\,.
\end{equation}
$H_\chi$ differs from $H_\phi$ since the canonical transformation $\phi \rightarrow \chi$ is time-dependent. The difference between the two Hamiltonians is given by the time derivative of the generating function of the canonical transformation~\cite{LLmeca}. In any case,   $H_\phi$ must be used to find the energy density on the right hand side of the Friedmann equation because this is the same energy associated to the energy-momentum tensor obtained by varying the matter Lagrangian with respect to the metric $g_{\mu \nu}$. 
Nevertheless, out of interest, we shall also compute the energy density associated to $H_\chi$ in our numerical analysis.

\subsubsection{Adiabatic subtraction}
In the previous paragraph we have defined the classical expression of the energy-momentum tensor $T_{\mu\nu}$. After quantization, one obtains the corresponding quantum operator $\hat{T}_{\mu \nu}$. 
The semi-classical approach consists in replacing the right-hand side of Einstein equation by the 
expectation value $\langle s \mid \hat{T}_{\mu \nu} \mid s \rangle$  where $\mid s \rangle$ is the 
quantum state of the scalar field. Without the exponential ultraviolet cutoff coming from non-commutativity, see Eq.\ \eqref{NCrho},
this expectation value is infinite and needs to be regularized. For example we have 
\begin{equation}
\langle 0 \mid \hat{H}_\chi(\eta) \mid 0 \rangle = \frac{1}{2} \delta^{(3)}(0) \int d^3 \textbf{k}  \left[ |\dot{v}_k(\eta)|^2 + \omega_k^2(\eta) |v_k(\eta)|^2 \right].
\end{equation}
This is the definition of energy density discussed in Ref.~\cite{mukhanov}.
The spatial energy density is obtained by dividing this expression by the 'volume' 
$(2\pi)^3a^4\delta^{(3)}(0)$. Performing the angular integral, the unrenormalized energy density becomes
\begin{equation}
\rho_\chi^{un} = \frac{1}{4\pi^2 a^4}  \int dk \, k^2  \left[ |\dot{v}_k(\eta)|^2 + \omega_k^2(\eta) |v_k(\eta)|^2 \right].
\end{equation}
From this quantity we have to subtract the (flat-space) vacuum contribution which is the same integral as above but with $v_k$ equal to the plane-wave expression, $v_k=\exp(-ik\eta)/\sqrt{2k}$.
This ``zero-point subtraction'' is not sufficient when the background is curved as the vacuum-subtracted expression, $\rho_{\chi, 0}$ is still divergent and one needs to resort to more sophisticated techniques, see e.\ g.\ \cite{BirDav}. There are several proposals in the literature on how to apply this regularization and on its domain of validity. In particular,  adiabatic subtraction has been recently discussed in relation to the regularization of the primordial power spectrum \cite{parker,adsub}. Usually,  one subtracts from the integrand of the divergent energy 
density its expansion in WKB, slowly varying, terms. The lowest order WKB term coincides with the ``instantaneous vacuum solution''. The subtraction needs to be performed up to the 4th WKB order to have a finite result in the ultraviolet (UV) regime. 

However, when the time dependence of the background is very strong, the different subtraction schemes  give very different, sometimes even negative results for the energy density and they cannot be trusted. In fact, when the time variation of the background is very rapid, the notions of particle and energy are simply not well defined. In our model, all the definitions proposed in the literature give the same result at late times, when the background is very slowly expanding,  and we can reliably compute the energy density at the end of inflation.

For the field $\phi$, the  energy density upon adiabatic subtraction in conformal time is (see appendix~\ref{adiabsub})
\begin{eqnarray}
{\rho}_{\phi}(\eta) &=& \frac{1}{4\pi^2a^4} \int dk \, k^2 \Bigg[  a^2\left| \left(\frac{v_k}{a}\right)^\bullet \right|^2 +k^2|v_k|^2+ \non \\ 
 &-& k -\frac{1}{2k}\left(\frac{\dot{a}}{a}\right)^2 -\frac{1}{8k^3}\frac{a\ddot{a}^2-2a\dot{a}a^{(3)}+4\dot{a}^2\ddot{a}}{a^3}+\mathcal{O}\left(\frac{\mathcal{H}^6}{k^5}\right)\Bigg]\ ,
\label{phienergy}
\end{eqnarray}
where the last three terms in the square bracket correspond to the counter-terms of adiabatic order 0 (the term $ -k$), 2 (the term $\propto 1/k$) and 4 (the term $\propto 1/k^3$). 
Both the 0th and 2nd order terms dominate in the UV, where we expect adiabaticity to be a valid approximation, so that there is a reasonable justification to subtract these terms. However, the 4th order term
is logarithmically divergent, hence it dominates also in the infrared (IR) regime, where adiabaticity is not at all verified.
In Ref.~\cite{parker}, the subtraction is justified with the fact that the final result converges. However, we still believe that it is not a good expression for the energy density of the particles generated during inflation \cite{adsub}. In particular, the 4th order subtraction introduces a new infrared divergence which is unphysical and which is not present in the un-subtracted result. Furthermore, as we shall see in our numerical results, this energy density can become negative, a most unphysical behavior for the energy density of a collection of free relativistic particles. 

In physical time, Eq.\ \eqref{phienergy} becomes
\bea
{\rho}_{\phi,4}(t) &=& \frac{1}{4\pi^2a^4} \int dk \, k^2 \Bigg[  a^4\left| \left(\frac{v_k}{a}\right)' \right|^2 +k^2|v_k|^2 + \non\\
 &-& k -\frac{1}{2k}a'^2 -\frac{1}{8k^3}(a^2a''^2+3a'^4-2a^2a'''a'-2a'^2a'')\Bigg].
\label{phienergyphys}
\eea

The notation $\rho_{\phi,n}$ indicates that we have performed the adiabatic subtraction up to the order $n$. By repeating the same calculations for $H_\chi$, we find, in conformal time
\bea
\rho_{\chi}(\eta)  &=& \frac{1}{4\pi^2 a^4} \int dk \, k^2 \, \Bigg[  |\dot{v}_k|^2 +\left(k^2-\frac{\ddot{a}}{a}\right)|v_k|^2 +\\\non 
&-& k +\frac{1}{2k}\frac{\ddot{a}}{a}+\frac{1}{8k^3}\left(\frac{\ddot{a}}{a}\right)^2 +\mathcal{O}\left(\frac{\mathcal{H}^6}{k^5}\right) \Bigg].
\label{chienergy}
\eea
Again, the three last terms in the square brackets correspond to the adiabatic counter-terms of order 0, 2 and 4. In physical time the above expression becomes
\bea
\rho_{\chi,4}(t) &=& \frac{1}{4\pi^2 a^4} \int dk \, k^2 \,\Big[  |a v_k'|^2 +\left(k^2-(a'^2+aa'')\right)|v_k|^2  +\non\\
 &-& k +\frac{1}{2k}(a'^2+aa'')+\frac{1}{8k^3}(a'^2+aa'')^2 \Big].
\label{chienergyphys}
\eea

As explained above, these definitions make sense only if the time dependence is weak, i.e. when the spectrum is
dominated by frequencies such that $k\gg \dot a/a$. In this situation, the adiabatic counter-terms are small and all the  expressions for the energy density do not differ by much. In opposition, in the infrared regime the adiabatic approximation breaks down and the result cannot be trusted.

In the context of the non-commutative regularization, the UV divergences  are cured by construction. Indeed, the exponential damping term in the integral measure  ensures the convergence of the energy density for $k\ra \infty$. Thus, to be fully consistent with the underlying non-commutative field theory, the above expressions for the energy density have to be modified in the UV where the exponential damping from non-commutativity sets in, see~\cite{maxinf}.  More precisely, eqs.~(\ref{phienergyphys}) and (\ref{chienergyphys})  have to be replaced by
\bea
 {\rho}_{\phi,4}(t) &=& \frac{1}{4\pi^2a^4} \int dk \, k^2 \exp\left[-2\theta \left(k\over a\right)^2\right]\Bigg[  a^4\left| \left(\frac{v_k}{a}\right)' \right|^2 +k^2|v_k|^2 +\non\\
 &-& k -\frac{1}{2k}a'^2 -\frac{1}{8k^3}(a^2a''^2+3a'^4-2a^2a'''a'-2a'^2a'')\Bigg],
\eea
and
\bea
\rho_{\chi,4}(t) &=& \frac{1}{4\pi^2 a^4} \int dk \, k^2 \exp\left[-2\theta \left(k\over a\right)^2\right] \,\Big[  |a v_k'|^2 +\left(k^2-(a'^2+aa'')\right)|v_k|^2 \non \\ 
&-& k +\frac{1}{2k}(a'^2+aa'')+\frac{1}{8k^3}(a'^2+aa'')^2 \Big].
\eea

\subsubsection{Analytical estimation}
\begin{figure}[tbp]
\centering
\includegraphics[width=0.5\textwidth]{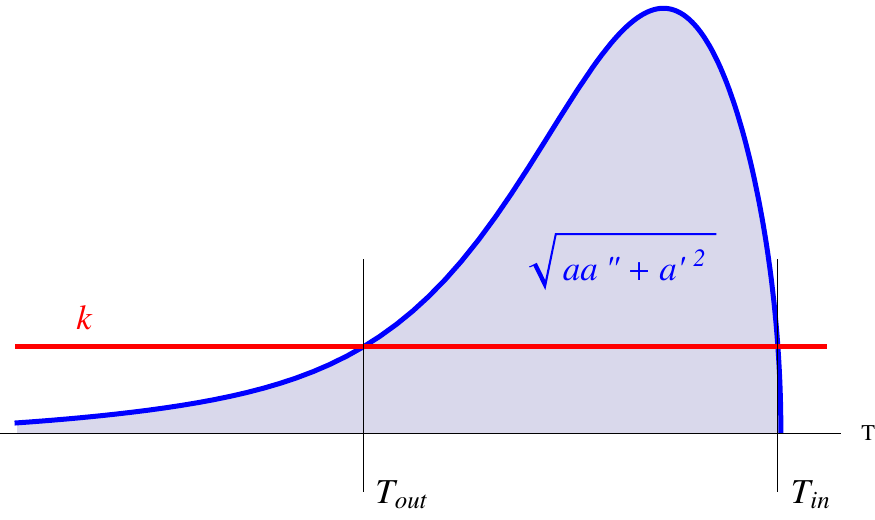}
\hfill 
        \caption{\label{f:Hubble} The dynamical mass $|m_\phi(T)|$ (blue line) and a co-moving wave number (red line) are indicated as functions of the dimensionless physical time $T$. When $k$ enters the shaded region the corresponding wavelength exits the Hubble scale, at $T_{\rm out}<0$, and it re-enters it when $k$ crosses the Hubble scale again at $T_{\rm in}>0$. At $T\sim 1$, $m_\phi^2$ becomes positive and the instability leading to particle creation is halted for all modes.} \end{figure}

Before presenting our numerical results, we perform an analytic approximation for the energy $\rho_\phi$. The un-subtracted, ultraviolet finite energy density is
\begin{equation}
\rho_{\phi}^{un}(\eta) = \frac{1}{4\pi^2a^4} \int dk \, k^2\exp\left[-2\theta \left(k\over a\right)^2\right] \left[  a^2\left| \left(\frac{v_k}{a}\right)^\bullet \right|^2 +k^2|v_k|^2 \right].
\end{equation}
Let us consider a mode with wave-number $k$ that originates as a sub-Hubble scale in the far past, crosses the Hubble horizon at $\eta_{\rm out}(k)$ and finally re-enters at $\eta_{\rm in}(k)$, see fig.~\ref{f:Hubble}.

As long as the mode is well inside the horizon, we have $\omega_k^2\equiv k^{2}-{\ddot a/a}\simeq k^{2}$, its amplitude can be approximated by a constant and the energy density can be set to zero.  After crossing the horizon, the mode starts to grow proportionally to the scale factor according to $(v_k/a)^\bullet=0$. To show this, note that in the super-horizon regime we have $\omega_k^2 \simeq - \ddot{a}/a$ and the equation of motion \eqref{vmodes} reduces to
\begin{equation}
\frac{\ddot{v}_k}{v_k}=\frac{\ddot{a}}{a},
\end{equation}
which has the general solution
\begin{equation}
v_k(\eta) = c_1(k)a(\eta) + c_2(k)a(\eta)\int_{\eta_0}^\eta {d\eta'\over a^2(\eta)}.
\end{equation}
Since the second term decays rapidly, we soon obtain $v_k \propto a$. In the computation of the energy density this leads to the approximation 
\bea
a^2\left|\left(\frac{v_k}{a}\right)^\bullet \right|^2 +k^2|v_k|^2 \simeq  \frac{k}{2}\frac{a(\eta)^2}{a(\eta_{\rm out}(k))^2}
\eea
that holds when the mode is super-horizon, namely for $\eta_{\rm out}(k)<\eta < \eta_{\rm in}(k)$. When $\eta > \eta_{\rm in}(k)$, the mode is again sub-horizon and thus of constant amplitude.  With this approximation, all modes with wave number above $k_{\rm max}\equiv\sqrt{\max(\ddot{a}/a)}$ are never excited and the  energy density is always finite. We then obtain the following approximation for $\rho_\phi$:
\begin{equation}
    \rho_{\rm est} = \frac{1}{4\pi^2a^4} \int dk \, k^2\exp\left[-2\theta \left(k\over a\right)^2\right] \left[ \frac{k}{2}f(k,\eta)^2-\frac{k}{2}\right],
\end{equation}
with 
 \begin{displaymath}
   f(k, \eta) = \left\{
     \begin{array}{lr}
       1 &  \eta < \eta_{\rm out}(k)\\
       a(\eta)/a(\eta_{\rm out}(k)) &  \eta_{\rm out}(k) < \eta < \eta_{\rm in}(k)\\
       a(\eta_{\rm in}(k))/a(\eta_{\rm out}(k)) & \eta > \eta_{\rm in}(k)
     \end{array}
   \right.
\end{displaymath} 
The term $-k/2$ in the integrand comes from the requirement that the energy density is zero in the far past and that it is continuous.

In the limit $a \rightarrow$ constant, the definitions $\rho_\phi$ and $\rho_\chi$ are the same and coincide with energy density of the field in Minkowski space. This is can be traced back to the fact that for $a=$constant, the canonical transformation is time-independent so the two Hamiltonian densities are the same.
In the opposite regime,  when $\mathcal{H}^2, \dot{\mathcal{H}} > k^2$, the adiabatic subtraction is no longer valid and and the counter-terms  diverge, causing  $\rho_\phi$ to become most likely negative. For 
$\rho_\chi$ the situation is even worse: the Hamiltonian $H_\chi$ has no minimum when $\omega_k^2<0$  which corresponds to the non-adiabatic regime~\cite{mukhanov}.  Negative energy densities are inevitable, with the formulae above, because a consistent  definition of particle and  energy  density is not known in strongly time-dependent backgrounds. However, as we shall see below, these expressions all coincide \emph{after} the inflationary phase, when $H\rightarrow 0$. Hence they can give a reasonable estimate of the final energy density of scalar particles. 

As soon as $\rho_{\phi}$ becomes of the order of the non-commutative energy density \eqref{e:rnc},
back-reaction becomes important and we can no longer trust our expansion law since the contribution from $\rho_\phi$ in the Friedmann equation cannot be neglected. The late non-commutative phase is altered by the energy of the particles produced in the early phase in such a way that the overall, final energy density can be different from the estimated one.

An advantage of the approximation $\rho_{\rm est}$ described above is that it is always positive and finite. However it may not be very precise, mainly because  the modes do not truly switch from the sub-horizon to the super-horizon regime instantaneously. In the non-commutative model at hand, the approximation 
is expected to be good when the time-dependence of the scale factor is very rapid or, equivalently, $\theta$ is very small as, in this case,  the transition between sub- and super-horizon regime  is very  sharp.

In the following section we compare the approximate value $\rho_{\rm est}$ against the numerical calculations of
$\rho_{\phi,0}$, $\rho_{\phi,2}$, $\rho_{\phi,4}$, $\rho_{\chi,0}$, $\rho_{\chi,2}$, $\rho_{\chi,4}$. 


\section{Numerical results}\label{nrsec}

\begin{figure}[ht]
\includegraphics[width=0.9\linewidth,angle=0]{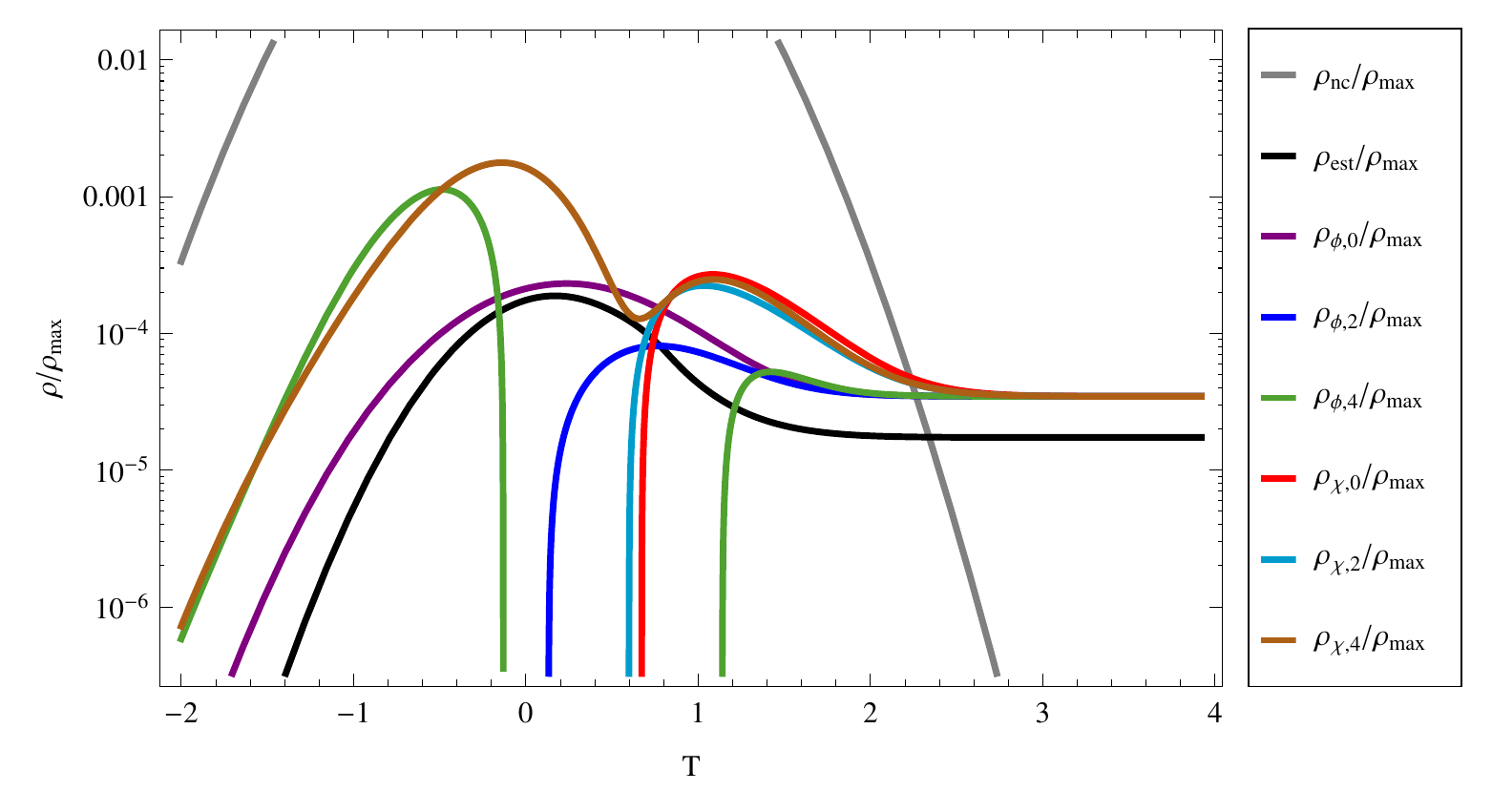} 
\caption{The different energy densities for $\alpha=10.6$ and $N_e=3$.}
\label{f:3ef}
\end{figure}
\begin{figure}[ht]
\includegraphics[width=0.9\linewidth,angle=0]{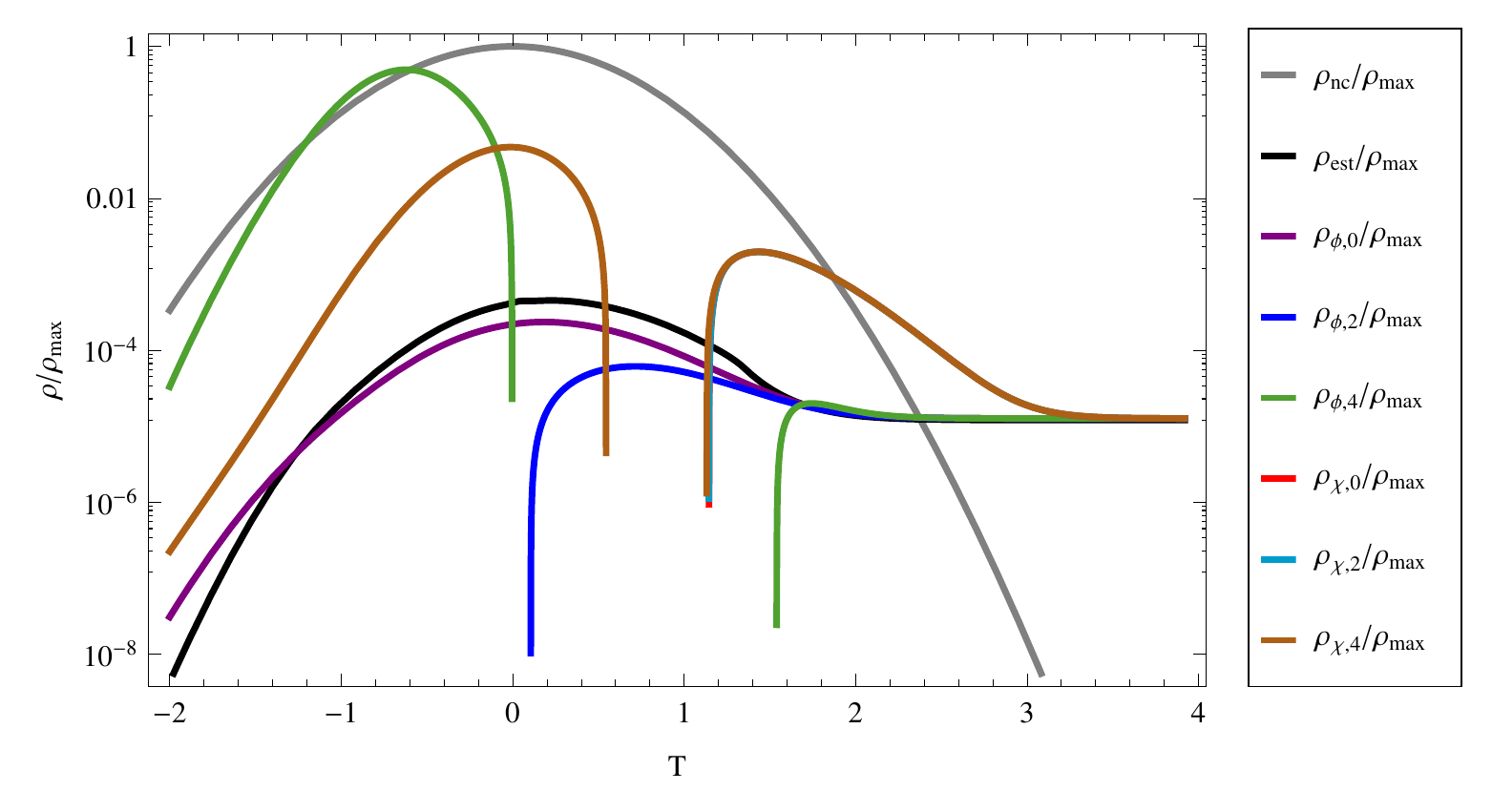} 
\caption{The different energy densities for $\alpha=10.6$ and $N_e=15$.}
\label{f:15ef}
\end{figure}
\begin{figure}[ht]
\includegraphics[width=0.9\linewidth,angle=0]{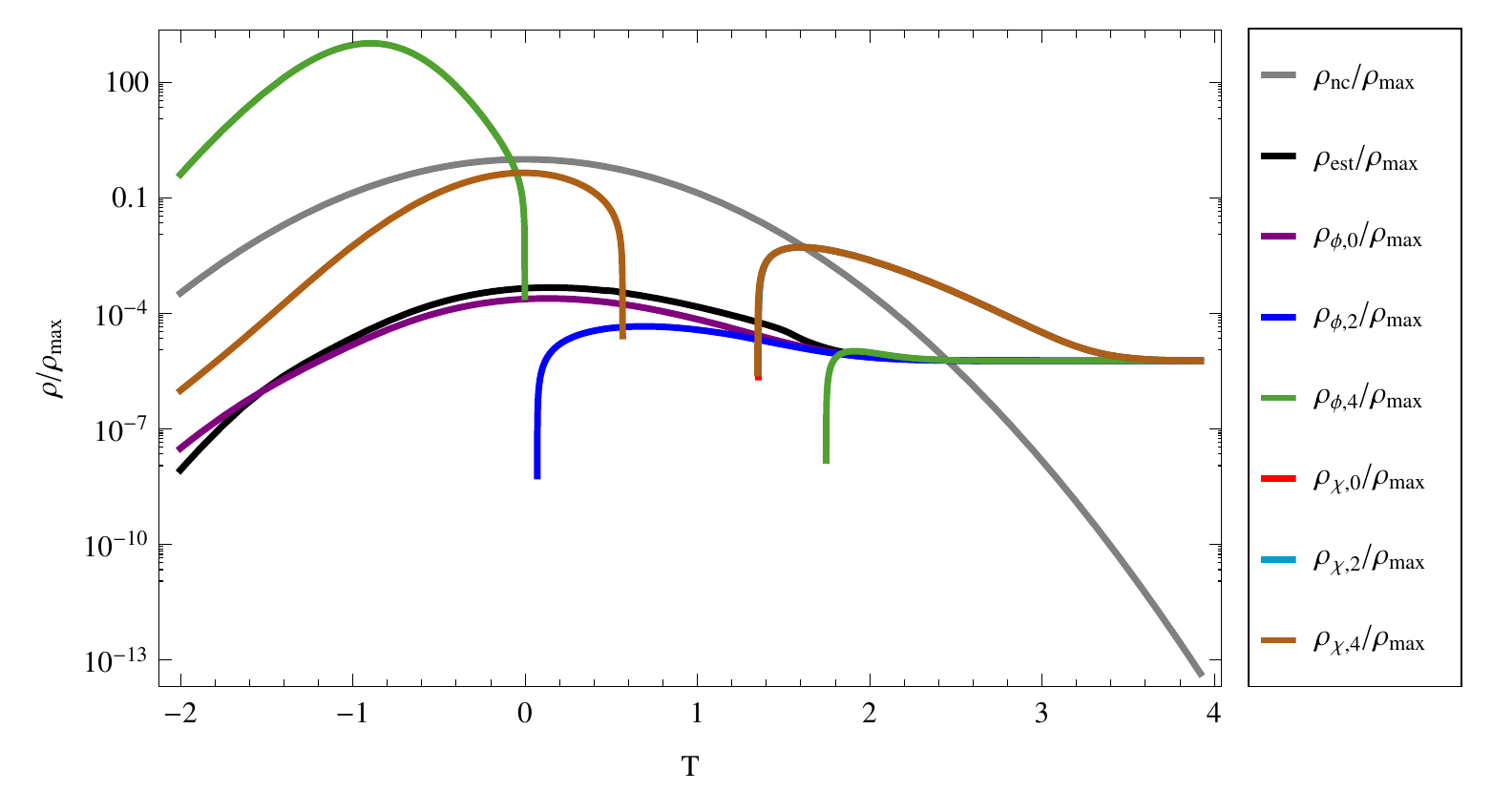} 
\caption{The different energy densities for $\alpha=10.6$ and $N_e=30$.}
\label{f:30ef}
\end{figure}

For large $\alpha\equiv\sqrt{\theta}/\ell_p$ and/or small $\beta=H_m\ell_p$ the value of $\rho_{\rm est}$ significantly underestimates the energy density both during and after the inflationary phase, but, in the opposite case, the situation improves a lot. In particular, if the relation (\ref{1scalarHtheta}) is strictly applied, such that 60 e-folds requires $\al=0.0136$, we expect the approximation to be excellent. In figs~\ref{f:3ef}, \ref{f:15ef} and \ref{f:30ef} we compare the various definitions of the energy density given in the text for $N_e=3,15$ and 30 respectively. The energy density, defined in  units of $\rho_{\max}=(32\pi^2\theta^{2})^{-1}(3/2)N_e^2\al^2$, is expressed as a function of the normalized cosmic time $T=t/\sqrt{8\theta}$. The grey line is the non-commutative energy density which drives inflation, defined in Eq.\ \eqref{e:rnc}. When the energy density of the particles generated by the inflationary expansion of the Universe becomes of the same order as $\rho_{\rm nc}$, it can no longer be neglected, inflation ends and the Universe becomes dominated by these massless particles. 

For rapid expansion, the numerical calculation becomes very difficult since the  result comes from the subtraction of $a^2|(v_k/a)^\bullet|^2 +k^2|v_k|^2$ and $k$ which, in the case, for example, of 60 e-folds, needs more than 100 decimals precision. Hence we have to compute $v_k$ with an accuracy better than 100 decimals. As long as the number of e-folds is small, this compensation is less severe, but already for $N_e=15$ the calculation takes several hours (with Mathematica and a precision of 30 digits).

Let us discuss fig.~\ref{f:3ef} in some detail. (For this case with $N_e=3$, we  also show the energy spectra in Appendix~\ref{a:spec}.) The energy densities $\rho_{\phi,n}$, (purple line, $n=0$, blue, $n=2$ and green $n=4$) as well as
$\rho_{\chi,n}$, are always significantly below $\rho_{\rm nc}$ (grey) for $T< 2$. The estimated energy density (black) underestimates the numerical results for all $T$. The gaps in the curves (or sharp raises/decays) correspond to the time intervals  when the energy density is negative (or to times where it becomes positive/negative). Both, $\rho_{\phi,2}$ and $\rho_{\phi,4}$  are negative when the time evolution is very rapid, i.e. around $T=0$ which  is very unphysical. Note also that at early times, $T<-0.5$, $\rho_{\phi,4}$ becomes  larger than $\rho_{\phi,0}$, another fact that shows how the subtraction scheme cannot be trusted in this regime. The maximum value $\rho_{\phi,4}$ at $T<0$ is dominated by the unphysical logarithmic infrared singularity introduced by the 4th order subtraction  (see spectra of $\rho_{\phi,4}$ and 
$\rho_{\chi,4}$  in Appendix~\ref{adiabsub}).

For $N_e=15$ and $N_e=30$ the situation is similar, just a bit more extreme, see 
figs.~\ref{f:15ef} and~\ref{f:30ef}. Now $\rho_{\phi, 2}$ remains negative until even later and the infrared divergence of $\rho_{\phi,4}$ becomes more prominent for $N_e=15$. For $N_e=30$,  it is even
much larger than $\rho_{nc}$ at early times, $T<-1$ when particle creation should still be very mild since expansion is still slow. For both cases $\rho_{\phi, 0}$ is well approximated by $\rho_{\rm est}$ after $T\simeq -1$. In both cases, all curves converge to the estimated energy density at 
$T\gtrsim 1.5$ for the $\rho_{\phi,n}$ and at $T>3.5$ for $\rho_{\chi,n}$. Therefore for  $N_e\gtrsim 15$ the final energy density is well approximated by $\rho_{\rm est}$.
  
 In terms of the dimensionless time $T$, the time interval of rapid evolution, i.e. inflation, always corresponds roughly to $-1.2<T<1.2$. 
Nevertheless, it is very difficult to calculate the energy density for large $N_e$ since it is hard to numerically obtain $v_k$ with the necessary very high precision which is given roughly by $(a_m/a(1))^4 \sim \exp(2N_e))$, if we consider that
$a^4\rho^{un}_{\phi}\theta^2$ is of order unity (see remark above). Nevertheless, all energy densities converge to the same value at late time where time evolution of the scale factor becomes slow. We find that this final energy density  
as a function of the non-commutative energy density is approximately given by

\be
\rho(t\rightarrow\infty) \simeq \rho_{\rm nc}(T=2.5)\,.
\ee

Note, however, that here we have computed the energy density of the created particles from one particle species,
while we typically need more than one species to have sufficient inflation, depending on the relation of $\beta$ and $\alpha$. With $\beta = \sqrt{N}/(\al^2\sqrt{12\pi})$ we can also rewrite eq.~(\ref{e:efolds}) as
\be
 N_e = \sqrt{\frac{2N}{3}}\frac{1}{\alpha} \,.
\ee

With $\al=10.6$, as in our case the true number of degrees of freedom is $N=(13N_e)^2$. The energy density of the created particles, in principle, has to be multiplied by this number, which raises 
$\rho_{\phi,0}$ above $\rho_{\rm nc}$ at the maximum, $T =0$ in the case $N_e=15$ and even earlier for $N_e=30$, but e.g.  $\rho_{\phi,2}$ remains below $\rho_{\rm nc}$ until $T\sim 1.5$ in both cases. 
Assuming that the true result would lie between $\rho_{\phi,0}$ and $\rho_{\phi,2}$, as a rule of thumb we can assume that radiation domination sets in at about $T\sim 1$ and therefore back-reaction reduces the effective number of e-folds by about 10\% (since Erf$(1) \sim 0.8$).

\begin{figure}[ht]
\begin{center}
\includegraphics[width=0.7\linewidth,angle=0]{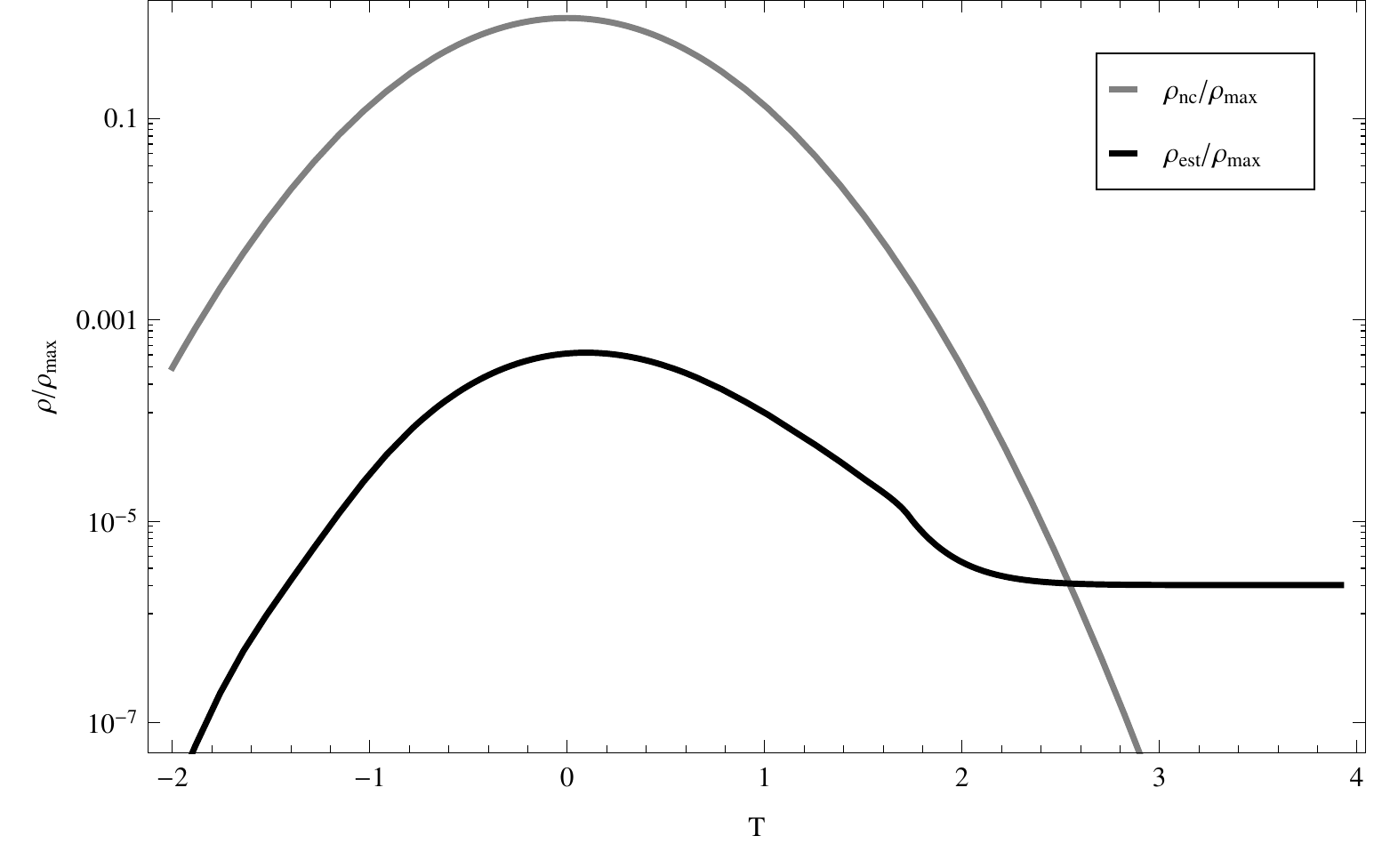} 
\caption{The non-commutative and the estimated energy densities for $\alpha=10.6$ and $N_e=60$.}
\label{f:beta}
\end{center}
\end{figure}

Even if we cannot present precise numerical results for an example with 60 e-folds of inflation, we
understand what happens qualitatively: for not very much fine tuned values of $\al$ and $\beta$ we can
obtain around 60 or more e-folds of inflation if we neglect particle creation. Including particle creation,  this number will be somewhat reduced but it is difficult to say by how much, since all our expressions for the energy density cannot be trusted in the very rapidly time-dependent phase. Nevertheless, the created particles, eventually dominate the energy density and terminate inflation. This model therefore has a natural, if not very graceful, exit from inflation. 

As an illustrative example, we have plotted the background energy density $\rho_{\rm nc}$ and the estimated energy density $\rho_{\rm est}$ for $\al= 10.6$  and $N_e =60$ in fig.~(\ref{f:beta}). The black curve has to be lifted by a factor $N\simeq 6\times 10^5$
in order to account for this number of degrees of freedom. This would lift it above the background density already before $T\sim 0$ and therefore reduce the number of e-folds by a factor 2. However, since our calculation of the energy density in quantum particle production cannot be trusted in this regime, and since the true maximum amplitude of the non-commutative energy density is not very well determined, 
this result is uncertain. If the truly  generated energy density remains below $\rho_{\rm nc}$, up to 
$T\gtrsim 1$ the model can provide close to 60 e-folds of inflation.

Nevertheless, these estimates tell us that this model has a natural maximum for the number of e-folds that can be achieved and which is probably somewhere around $N_e\sim 50$ -- $60$.


\section{Conclusion}\label{consec}


In this paper we have considered a model of non-commutative inflation proposed in Ref.~\cite{maxinf}.
We have found that this model leads to a rapid burst of particle creation which we term ``explosive particle creation''. This burst eventually terminates inflation and it is not certain whether sufficient inflation can actually be achieved. 

However, if this is possible, particle creation provides an exit mechanism from inflation. 
Inflation terminates at $T_{\rm end}\sim 1$ when the energy density of the generated relativistic particles takes over. After this, the Universe is radiation-dominated with an energy density given by
\be 
\rho =\rho(T_{\rm end})\left(\frac{a(T_{\rm end})}{a}\right)^4\,.
\ee
We expect interactions to be efficient so that these particles  rapidly thermalize and we end up with a hot plasma of elementary particles as in standard cosmology.

We conclude with a comment on the power spectrum of primordial fluctuations. It can be shown that it strongly depends on the adiabatic subtraction scheme. If we subtract up to the second adiabatic order, the resulting power spectrum is quite flat, $\rho(k) \propto \log(k)$, in the UV regime. Therefore, in addition to providing the background radiation energy density at the end of non-commutative inflation, this model might also lead to a  nearly scale invariant spectrum of large scale cosmological density fluctuations. However, to compute their amplitude and the spectrum  in detail, we should push the calculations in the far infrared, where adiabatic subtraction is not justified. How to overcome this difficulty remains an open problem for a future project.


\acknowledgments

H.\ P.\ and R.\ D.\ are supported by the Swiss National Science Foundation. M.\ R.\  is supported by a grant of the ARC 11/15-040 convention.


\appendix

\section{Adiabatic subtraction}\label{adiabsub}

In this appendix we briefly summarize the adiabatic subtraction scheme, for more details see \cite{BirDav}.
We assume that the modes can be written in the WKB form
\begin{equation}
v_k^{ad}=\frac{1}{\sqrt{2\Omega_k}}e^{-i\int \Omega_k d\eta}.
\end{equation}
Plugging this expression into the equation of motion \eqref{vmodes}, one finds that $\Omega_k$  satisfies
\begin{equation}
\Omega_k^2=k^2-\frac{\ddot{a}}{a}+\frac{3\dot{\Omega}_k^2}{4\Omega_k^2}-\frac{\ddot{\Omega}_k}{2\Omega_k}.
\label{Omega}
\end{equation}
One then solves this equation iteratively to the order $n$, which concides with the number of derivatives of $\Omega_k$. This procedure is valid only in the adiabatic regime, i.\ e.\ for $k^2 \gg \mathcal{H}^2, \dot{\mathcal{H}} \sim 
(\dot\Omega_k/\Omega_k)^2$. The term $\ddot{a}/a$ is to be considered as order two (see \cite{adsub}). We denote the expansion to order $n$ of $\Omega_k$ and of the mode function $v_{k}$ by $\Omega_{k,n}$ and  
\bea
v_k^{ad,n}=\frac{1}{\sqrt{2\Omega_{k,n}}}e^{-i\int \Omega_{k,n} d\eta}
\eea
 respectively.

At zero order, we simply have $\Omega_{k,0} = k$ and  $v_k^{ad,0}=v_k^{vac}$.  Removing this contribution is equivalent to removing the Minkowski, zero-point  vacuum contribution. To find the second order, we insert  $\Omega_{k,0}$ on the left-hand side of equation \eqref{Omega} and expand in $\ddot a/(ak^2)\ll 1$ yielding
\begin{equation}
\Omega_{k,2}=k\left(1-\frac{\ddot{a}}{2ak^2}\right).
\end{equation}
By repeating the operation, we finally obtain the fourth order expression for $\Omega_k$,
\begin{equation}
\Omega_{k,4}= k \left(1-\frac{\ddot{a}}{2k^2a}+\frac{a^{(4)}a^2-2a\ddot{a}^2-2a^{(3)}a\dot{a}+2\dot{a}^2\ddot{a}}{8k^4a^3}\right).
\end{equation}
As said before, these expressions are only valid in the adiabatic regime, although some authors argue that the subtraction can be employed even when the expansion is strongly time-dependent~\cite{parker}.

Instead of solving the mode equation for $v_k$ one also could solve the equivalent equation
for the amplitude $M_k =1/(2\Omega_k)$ which follows from \eqref{Omega}. One might think that this is numerically less involved since one does not have to keep track of  the oscillations of the modes $v_k$, see~\cite{1207.3227}. However, to achieve the very high accuracy needed for our purposes, we have found that the numerical problem is as hard as the one for $v_k$. 
\newpage

\section{Contribution of modes to the energy density}\label{a:spec}
The energy density spectra, $\rho^{-1}_{\max}(d\rho/d\log K)$ are shown as functions of the dimensionless variables 
$T$ and $K$. All of them, except $\rho_{\phi, 0}$ do have regions where they are negative.
\begin{figure}[ht!]
\begin{center}$
\begin{array}{cc}
\includegraphics[width=0.45\linewidth,angle=0]{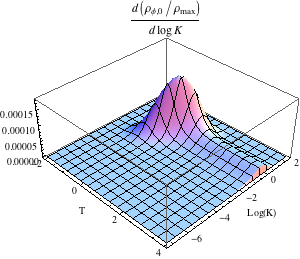} &
\includegraphics[width=0.45\linewidth,angle=0]{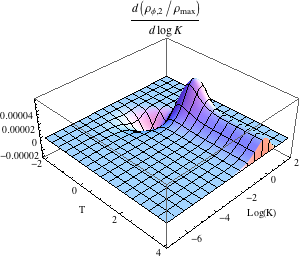} \\ 
\includegraphics[width=0.45\linewidth,angle=0]{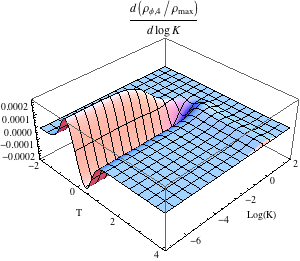} &
\includegraphics[width=0.45\linewidth,angle=0]{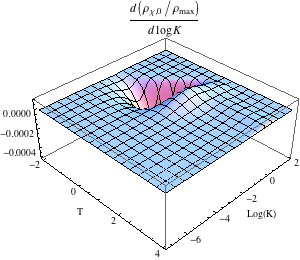} \\
\includegraphics[width=0.45\linewidth,angle=0]{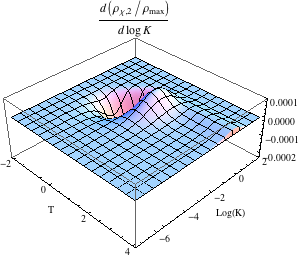} &
\includegraphics[width=0.45\linewidth,angle=0]{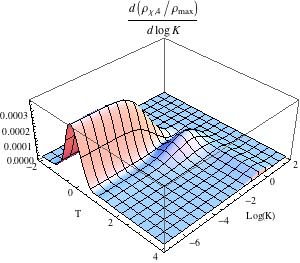} \\
\end{array}$
\end{center}
\caption{Contribution to the energy density as function of K and T for $N_e=3$}
\label{energy3d}
\end{figure}

\newpage



\end{document}